\documentclass[aps,prl,twocolumn,superscriptaddress,nobibnotes,amsmath,amssymb]{revtex4-2}
\usepackage{graphicx}
\usepackage{color}
\usepackage[explicit]{titlesec}
\usepackage[normalem]{ulem}
\usepackage{gensymb}
\usepackage{hyperref}
\hypersetup{
	colorlinks=true, % Set to false to disable coloring links
	citecolor=magenta, % The color of citations
	linkcolor=magenta, % The color of references to document elements (sections, figures, etc)
	urlcolor=magenta, % The color of hyperlinks (URLs)
}

\newcommand{\myvareq}[4]{$#1_{\text{#2}}=#3\,\mathrm{#4}$}

\newcommand{\myq}[2]{$#1\,\mathrm{#2}$}
\newcommand{\myqapprox}[2]{$\approx #1\,\mathrm{#2}$}

\newcommand{\vg}[0]{$V_{\text{g}}$ }
\newcommand{\vgeq}[1]{$V_{\text{g}}=#1\,\text{V}$ }
\newcommand{\vgg}[1]{$V_{\text{g}}>#1\,\text{V}$ }
\newcommand{\vgl}[1]{$V_{\text{g}}<#1\,\text{V}$ }
\newcommand{\vsd}[0]{$V_{\text{sd}}$ }

\newcommand{\vthapprox}[1]{$V_{\text{th}} \approx #1\,\mu V$}

\begin{document}
\title{Localization of the Helical Edge States in the Absense of External Magnetic Field}
\author{A.V.~Bubis}
\affiliation{Skolkovo Institute of Science and Technology, Nobel str. 3, 143026, Moscow, Russian Federation}
\affiliation{Institute of Solid State Physics, Russian Academy of Sciences, 142432 Chernogolovka, Russian Federation}
\author{N.N.~Mikhailov}
\affiliation{Institute of Semiconductor Physics, Novosibirsk 630090, Russian Federation}
\affiliation{Novosibirsk State University, Novosibirsk 630090, Russian Federation}
\author{S.A.~Dvoretsky}
\affiliation{Institute of Semiconductor Physics, Novosibirsk 630090, Russian Federation}
\author{A.G.~Nasibulin}
\affiliation{Skolkovo Institute of Science and Technology, Nobel str. 3, 143026, Moscow, Russian Federation}
\affiliation{Aalto University, P. O. Box 16100, 00076 Aalto, Finland}
\author{E.S.~Tikhonov}
\email[e-mail:]{tikhonov@issp.ac.ru}
\affiliation{Institute of Solid State Physics, Russian Academy of Sciences, 142432 Chernogolovka, Russian Federation}
\begin{abstract}
Theoretically, the helical edge states of two-dimensional topological insulators are protected from coherent backscattering due to nonmagnetic disorder provided electron interactions are not too strong. Experimentally, the edges typically do not demonstrate the systematic and robust quantization, at the same time little is known about the sub-Kelvin temperature behavior. Here, we report the surprising localization of the edge states in an \myq{8}{nm} HgTe quantum well in zero magnetic field at millikelvin temperatures. Additionally, the magnetoresistance data at \myq{0.5}{K} for the edges few micrometers long suggests the field-dependent localization length 
$l_B\propto B^{-\alpha}$, with $\alpha$ ranging approximately from $1.6$ to $2.8$ at fields $B\lesssim0.1\,\text{T}$ and $\alpha\approx1.1$
at higher fields up to~$0.5\,\text{T}$. In the frame of disordered interacting edge, these values of $\alpha$ correspond to the Luttinger liquid parameters \mbox{$K\approx 0.9-1.1$} and \mbox{$K\approx 0.6$}, respectively. We discuss possible scenarios which could result in the zero magnetic field localization.
\end{abstract}
\maketitle
The concept of the quantum spin Hall (QSH) effect~\cite{RevModPhys.82.3045,RevModPhys.83.1057} is manifested in the existence of gapped bulk insulators with the edge conduction due to helical electrons for which the spin and the momentum are locked. 
Time-reversal symmetry (TRS) protects edge channels of these two-dimensional topological insulators (2D TIs) against single-particle coherent backscattering. From here, in the experiment at low enough temperature~$T$ one could have naively expected the quantized edge conductance~$G_q=e^2/h$.
Still, it is common that at low~$T$ one usually observes almost absent or only very weak $T$-dependence for the edge conductances $G\ll G_q$~\cite{PhysRevB.89.125305}. Such a behavior is believed to be due to yet unexplained phase breaking mechanisms but may be phenomenologically captured, e.g., in the model of the conducting charge puddles in the insulating bulk of the 2D TI~\cite{PhysRevB.94.045425}. 

Beyond the single-particle description, it was realized early on that the picture of ideal helical edge states can be significantly modified by two-particle scattering processes~\cite{PhysRevLett.96.106401,PhysRevB.73.045322}.
The strength of the omnipresent inter-electron Coulomb interaction is usually expressed in terms of the Luttinger liquid parameter~$K$ which is also dependent on the Fermi velocity~$v_{\text{F}}$ and the system geometry~\cite{PhysRevB.79.235321,PhysRevLett.102.096806}. The value of~$K$, with $0<K<1$ corresponding to repulsive interactions and $K=1$ corresponding to the noninteracting electrons, crucially defines the transport properties of the 2D TI~\cite{dolcetto2016}. As an example, the recent theoretical paper~\cite{PhysRevB.98.054205} studies the phase diagram of a disordered interacting edge and predicts the possibility of an insulating edge for strong enough interactions~$K<3/8$. This is in line with early works~\cite{PhysRevLett.96.106401,PhysRevB.73.045322} predicting the possibility of localization for strong enough interactions.

So far, only a few experimental studies addressed the measurements of~$K$ in the helical edge channels~\cite{PhysRevLett.115.136804,Sthler2019}. The interpretation of the temperature dependence of the conductance or the bias dependence of the differential conductance in a transport experiment may be ambiguous since the extracted value of~$K$ depends on the underlying theory~\cite{PhysRevLett.115.136804,PhysRevB.93.241301}. This approach is further complicated by the fact that above~\myq{1}{K} $R(T)$-behavior is usually weak~\cite{PhysRevB.89.125305} while the data at sub-Kelvin temperatures are scarce and poorly consistent with each other~\cite{Konig2007a,PhysRevLett.114.096802,PhysRevLett.114.126802,PhysRevLett.123.056801}. At the same time, the tunneling spectroscopy approach of~\cite{Sthler2019} is not applicable to the most studied HgTe/CdTe and InAs/GaSb realizations of 2D~TIs. We also note that the proposed corner junction experiments on the interedge tunneling~\cite{PhysRevLett.102.076602,PhysRevLett.107.096602} have not yet been realized, the notable recent exception being the experiment in the standard quantum point contact geometry~\cite{Strunz2019}. Furthermore, experimentally the situation may be complicated by the possible presence of magnetic impurities~\cite{PhysRevLett.111.086401}, Rashba spin-orbit coupling originating from the electric field of the gate electrode~\cite{PhysRevLett.104.256804}, or hyperfine interaction with the nuclear spins~\cite{PhysRevB.96.081405}.

Since TRS is an essential ingredient for the topological protection, external magnetic field is an important knob to provide additional information. Introduction of the magnetic field opens a Zeeman gap in the edge spectrum~\cite{Wu2018} and couples the counterpropagating edge modes. As a result, the restored coherent backscattering must decrease the conductance or even localize the edges, provided the phase coherence is preserved. This common anticipation, while partly confirmed already in the very first experimental paper, was later on contrasted with the observations of only weakly $B$-dependent transport in both HgTe and InAs/GaSb quantum wells (QWs)~\cite{Ma2015,PhysRevLett.114.096802}. Theoretically, this robustness of the edge-state transport to the magnetic fields as high as several Tesla was later attributed to the Dirac point being hidden in the bulk band rather than in the bulk gap~\cite{PhysRevB.97.045420,PhysRevB.98.201404}. On the other hand, localization was recently reported for the helical edges in $d=8$ and \myq{14}{nm} HgTe QWs with drastic changes in conductance for the \myvareq{d}{}{8}{nm} case already at few mT perpendicular fields~\cite{PhysRevLett.123.056801}. With similar behavior of both quasi-ballistic and resistive edges in zero $B$-field down to the lowest temperatures, this experiment demonstrated that the edge transport with $G\ll G_q$ is in fact due to the topological protection. Unfortunately, one has to admit no clear picture of the magnetic-field-dependent localization of the helical edge states exists at the moment.

\begin{figure}[h]
\begin{center}
\vspace{0mm}
\includegraphics[width=86mm]{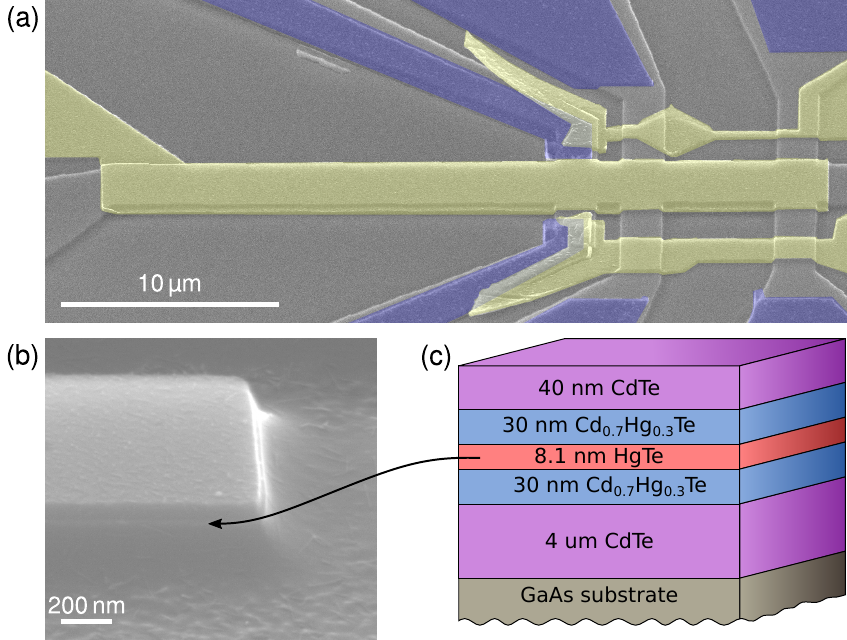}
\end{center}
\caption{(a) False color tilted SEM micrograph of the device~D2. Three galvanically isolated top gate electrodes are marked with yellow. The evaporated aluminum contacts are colored in blue. 
(b)~The mesa after wet etching. The \myq{8}{nm} wide QW is visible as a white narrow line. 
(c)~Schematic view of the heterostructure before the lithography. The helical edge states exist along the boundary of the HgTe layer~(red).}
\label{fig1}
\end{figure}

In the present work, we demonstrate the helical edges may localize at millikelvin temperatures even in zero magnetic field and discuss the possible scenarios of localization. 
Our samples are based on the \myq{8}{nm} wide \mbox{HgTe/CdHgTe} QW grown by molecular beam epitaxy, see Fig.~\ref{fig1}. For patterning of the mesa we used the approach of~\cite{Bendias2018} which results in an uncontaminated mesa sidewall, thus the thin HgTe layer can be observed in SEM, see the white narrow line in Fig.~\ref{fig1}(b). Below we present the data from two geometrically similar devices~D1 and~D2 etched simultaneously. Due to technical difficulties, we were not able to cool down device~D1 to millikelvin temperatures. The measurements at~\myq{50}{mK} were performed in a BlueFors-LD250 dilution refrigerator not equipped with a solenoid. The measurements at~\myq{4.2}{K} and~\myq{0.5}{K} were performed in a $^3$He insert with the external magnetic field perpendicular to the QW plane. The details of DC lines filtering and device fabrication may be found in Supplemental Material.

\begin{figure}[h]
\begin{center}
\vspace{0mm}
\includegraphics[width=3.4in]{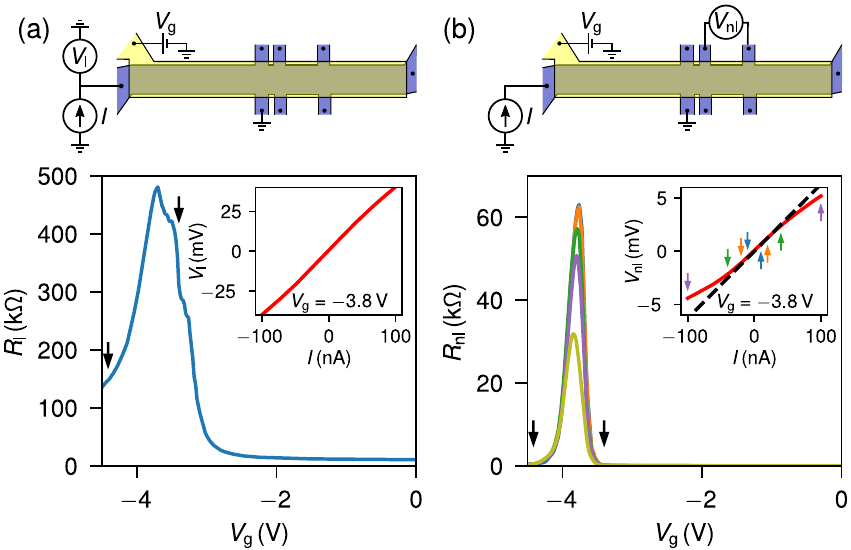}
\end{center}
\caption{
Nonlocal transport measurements at~\myvareq{T}{}{4.2}{K}. (a)~Local two-terminal linear-response resistance of the \myq{20}{\mu m} edge on device~D1. The inset demonstrates the local $I$-$V_{\text{l}}$~characteristics of the edge at the CNP. (b)~Nonlocal resistance measured in the configuration shown above for $I=10$, $20$, $40$, $100$ and \myq{300}{nA} (different colors, note the blue and orange curves are almost indistinguishable). The inset demonstrates the nonlocal $I$-$V_{\text{nl}}$~characteristics.}
\label{fig2}
\end{figure}

In our devices, the QSH regime is realized by tuning the Fermi energy level to the bulk energy gap using the voltage~$V_{\text{g}}$ applied to the gate electrode. At low temperatures, the devices demonstrate typical dependencies of resistance on the gate voltage~$V_{\text{g}}$. For the case of the \myq{20}{\mu m} edge, see Fig.~\ref{fig2}(a), gate voltages \vgg{-3.4} correspond to the n-type conduction in the bulk, gate voltages \vgl{-4.5} correspond to the p-type conduction in the bulk, while the intermediate range of~$V_{\text{g}}$ corresponds to the case when the Fermi level is inside the insulating bulk. In this region of~$V_{\text{g}}$, the so-called charge neutrality point region (CNP), the conduction through the device is dominated by the edge states which is manifested as a maximum in the dependence $R(V_{\text{g}})$. Note that the exact range of gate voltages corresponding to the~CNP may differ for different devices. In the CNP, the $I$-\vsd curves are close to linear, see the inset of Fig.~\ref{fig2}(a). 
We note that for the edges not longer than few micrometers we generally do not see any systematic dependence of resistance~$R$ on the edge length~$L$ with the values of resistance at the CNP on the order of \myq{100}{k\Omega} at~\myq{4.2}{K}. For the longer edges the dependencies~$R(L)$ are monotonously growing. 

We verify the edge transport at the CNP using the voltage measurements on the contacts seemingly not lying on the current path~\cite{Roth2009,PhysRevB.84.121302,tikhonov2015,Liu2019}, see Supplemental Material Fig.~1. 
At the CNP, the voltage along the edge of the device monotonously falls when moving from the biased contact to the ground reflecting the negligible bulk transport. One can estimate the bias voltages on the edges for which the bulk shunt may be safely disregarded as we now demonstrate for the \myq{20}{\mu m} edge. Fig.~\ref{fig2}(b) demonstrates the nonlocal resistance $R_{\text{nl}}=V_{\text{nl}}/I$ measured in the configuration shown above for five different bias currents $I=10$, $20$, $40$, $100$ and \myq{300}{nA}. In terms of~$V_{\text{g}}$, the peak of the nonlocal signal coincides with the peak in the two-terminal resistance, and the nonzero nonlocal voltage allows one to identify the experimentally relevant range of~$V_{\text{g}}$, see the black arrows. For $I\lesssim40\,\text{nA}$, $R_{\text{nl}}$ is current-independent which is manifested in the linear $I$-$V_{\text{nl}}$ curve, see the inset of Fig.~\ref{fig2}(b). At further increasing~$I$, $R_{\text{nl}}$ monotonously falls, see the green, violet and olive curves in Fig.~\ref{fig2}(b) and $I$-$V_{\text{nl}}$ curve starts demonstrating the sublinear shape indicating the opening current leak to ground through the bulk of the device. For the \myq{20}{\mu m} edge, \myvareq{I}{}{40}{nA} corresponds to $V_\mathrm{l}$~\myqapprox{15}{mV} which is comparable to the expected value of the bulk gap $\Delta_{\text{bulk}}\approx30\,\text{meV}$ and allows one to exclude the contribution of the bulk transport at smaller biases. 

\begin{figure}[t]
\begin{center}
\vspace{0mm}
\includegraphics[width=3.4in]{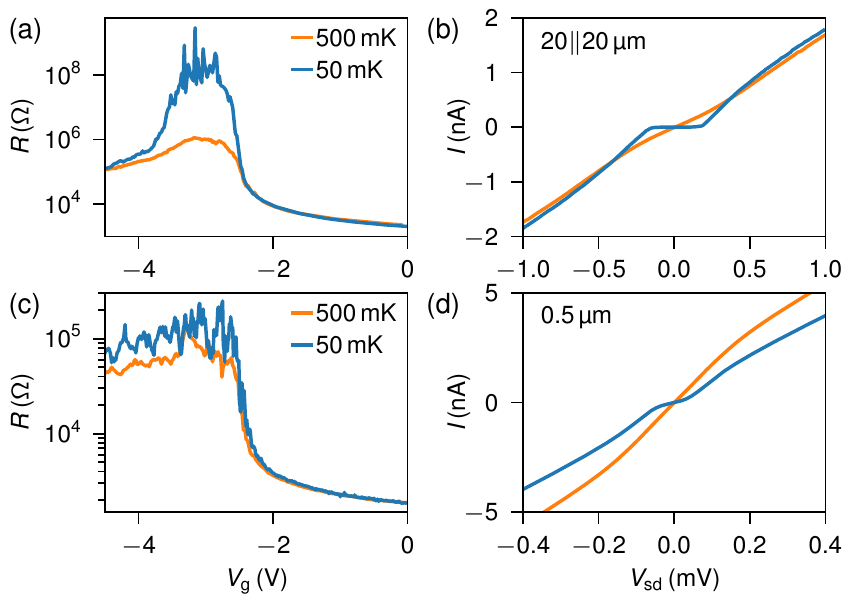}
\end{center}
\caption{Localization of the edge states at lowering the temperature on the device~D2. 
(a,~b) Two-terminal linear-response resistance as a function of $V_{\text{g}}$ and the $I$-\vsd curve at the CNP for the \myq{20}{\mu m} edge at \myvareq{T}{}{0.5}{K} (orange) and \myq{50}{mK} (blue). Lowering the temperature leads to the substantial increase of resistance at the CNP accompanied by the development of significant mesoscopic fluctuations. The $I$-\vsd curves are close to linear at \myvareq{T}{}{0.5}{K} but demonstrate the gap-like feature at \myvareq{T}{}{50}{mK}. (c,~d) Similar data for the \myq{0.5}{\mu m} edge.
}
\label{fig3}
\end{figure}

We continue our discussion with the temperature dependence of resistance of the helical edge states at sub-Kelvin temperatures in zero magnetic field on the device~D2. Hereinafter, we label the straight edges of the device~D2 using their length and label the several hundred nm corner-shaped edge as ``corner'' (see Supplemental Material Fig.~2 for the details).
For the \myq{20}{\mu m} edge, see Fig.~\ref{fig3}(a), both at \vgg{-2.4} and at \vgl{-3.6} lowering the temperature from \myvareq{T}{1}{0.5}{K} down to \myvareq{T}{2}{50}{mK} only weakly influences the value of resistance reflecting the metallic type conduction of the two-dimensional electron and hole gases. Conduction through the edge, however, demonstrates strong dielectric behavior with emerging giant mesoscopic fluctuations. Here, the increase of resistance at lowering the temperature is greater than two orders of magnitude. Assuming the activation-like behavior, one can estimate the value for the activation energy as $\Delta(20\,\mathrm{\mu m})\approx k_{\text{B}}T_2\ln(R_2/R_1)\approx20\,\mathrm{\mu eV}$. Quantitatively, the discussed effect is most prominent for the long edges but it persists even for the shortest ones, see panel~(c) for the case of~\myq{0.5}{\mu m} edge. Supplemental Material Fig.~3 provides more $R(V_{\text{g}})$ curves for the other edges as well as the $T$-dependence of the mean linear response conductance (averaged in the \myq{0.3}{V} range of $V_{\text{g}}$ in the CNP region) for the \myq{5}{\mu m}~edge with the estimate for the activation energy~$\Delta(5\,\mathrm{\mu m})\approx16\,\mathrm{\mu eV}$.

The $I$-\vsd characteristics in zero magnetic field at lowest~$T$ become strongly nonlinear with the threshold-like behavior.
Fig.~\ref{fig3}(b) compares the $I$-\vsd characteristics measured at two temperatures at the CNP for the \myq{20}{\mu m} edge. While at \myvareq{T}{}{0.5}{K} the curve is almost linear (see Supplemental Material Fig.~4 for the differential resistance curves at $B=0$ in the CNP region), at \myvareq{T}{}{50}{mK} it demonstrates the gap-like feature with the threshold value of \vthapprox{200}. We note that the corresponding energy scale is by far smaller than the bulk energy gap~$\Delta_{\text{bulk}}$ of the \myq{8}{nm} HgTe QWs. We also note that $eV_{\text{th}}$ is considerably greater than the estimate for the activation energy~$\Delta(20\,\mathrm{\mu m})$. This indicates that the applied bias is shared among several strongly localized electronic states along the edge~\cite{PhysRevLett.123.056801}. For the shorter, \myq{0.5}{\mu m} edge, the $I$-\vsd characteristics is qualitatively analogous, however, with a less pronounced gap-like feature and a smaller energy scale, see Fig.~\ref{fig3}(d). Altogether, the observations of Fig.~\ref{fig3} demonstrate the localization of the edge states at lowering temperature. 

To the best of our knowledge, this is the first observation of localization of the helical edge states in a 2D~TI in zero magnetic field. It is in glaring contrast with the sub-Kelvin measurements of~\cite{PhysRevLett.123.056801} where down to \myq{50}{mK} the edges almost didn't demonstrate any $T$-dependence of conductance. We pay attention to the difference in fabrication of the devices. Here, for mesa formation we used e-beam lithography followed by wet etching, while the authors of~\cite{PhysRevLett.123.056801} used dry Ar plasma etching applied to photolithographically defined pattern. Yet another difference is in the way the ohmic contacts are realized -- here, we evaporate Ti/Al contacts after preliminary cap removal with the Ar gun, while in~\cite{PhysRevLett.123.056801} the contacts were indium soldered. Unfortunately, we are not sure how exactly the processing might have led to the observed drastic difference in the low-$T$ behavior.
 
We now discuss the potential scenarios of localization of the helical edge states in the absence of magnetic field. The first scenario could have relied on the presence of anisotropic magnetic disorder~\cite{PhysRevLett.111.086401, PhysRevB.93.241301, Kurilovich2017}. For the case of spin $S=1/2$ impurities, one can crudely estimate the number of impurities that would account for the conductance decrease on the order of $e^2/h$, as $N_{\text{imp}}\sim4\xi^4M^2/J^2\approx5000$, 
where $J\approx0.1\,\text{eV\,nm}^2$ is the anisotropic exchange coupling, $|M|=30\,\text{meV}$ is the band gap, and $\xi\approx10\,\text{nm}$ is the characteristic width of the edge states~\cite{Kurilovich2017}. For the $8\,\text{nm}\times10\,\text{nm}\times1\,\mathrm{\mu m}$ edge with $10^6$~atoms and resistance of $100\,\mathrm{k\Omega}$, this number of impurities seems improbable given the QW is grown from $99.9999$\% pure material. This estimate indicates that magnetic impurities on their own do not dominate the edge resistance and are hardly the reason for localization. We note, however, that for the case of strong enough electron-electron interactions in the edge even a single magnetic impurity can lead to insulating behavior.~\cite{Maciejko2009}

The nuclear spins are also known to suppress the conductance of the 2D~TI edges at low temperatures provided electron-electron interactions are strong~\cite{PhysRevB.96.081405}. Our QWs are grown from the naturally abundant Hg and Te atoms, approximately $19$\% of which have a nonzero nuclear spin. In this case, the localization temperature may indeed fall in the millikelvin range, however the localization length is expected to be as high as several millimeters~\cite{PhysRevB.97.125432}. Thus, the hyperfine interaction is also unlikely the reason for localization in our micrometer long edges. 

Yet another possibility for the observations of Fig.~\ref{fig3} is the interaction-driven localization of the disordered edges~\cite{PhysRevLett.96.106401,PhysRevB.73.045322,PhysRevB.98.054205,PhysRevB.90.075118}. In this picture, the exact value of the Luttinger parameter~$K$ determines not only if the localization occurs but it also defines the magnetotransport behavior. In particular, the typical edge conductance $G_{\text{typ}}=\exp(\overline{\ln G})$ in the magnetic field~$B$ is expected to obey $G_{\text{typ}}\propto \exp\left(-l/l_{\text{B}}\right),$
where 
the field-dependent localization length follows the power law $l_B\propto B^{-2/(3-2K)}$. 
We perform magnetotransport measurements at~\myvareq{T}{}{0.5}{K} to see if the data fit this prediction. On the one hand, this temperature is high enough so that the discussed above low-$T$ localization is not yet developed. On the other hand, we checked that at higher temperature \myq{4.2}{K} the edges generally do not display any response to the external magnetic field at least up to~\myq{1.5}{T}, see Supplemental Material Fig.~5. We concentrate on the relatively short edges to avoid the possible size-effects, see Supplemental Material Fig.~6 for the results concerning the longer edges. 

\begin{figure}[h]
\begin{center}
\vspace{0mm}
\includegraphics[width=3.4in]{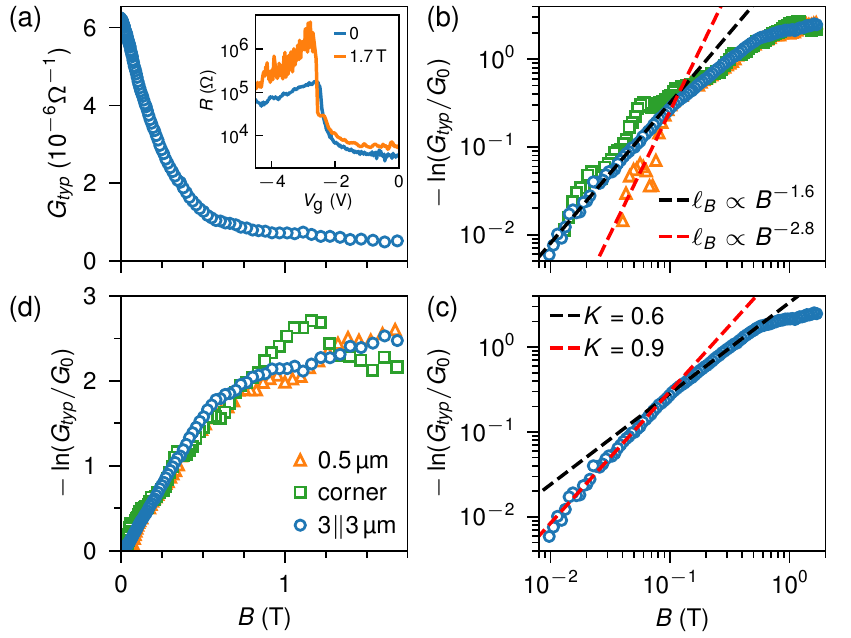}
\end{center}
\caption{Influence of the external magnetic field on the edge conductance at~\myvareq{T}{}{0.5}{K}. (a)~Typical conductance of the $3\|3\,\mathrm{\mu m}$ edge measured in the \myq{20}{mV} range of the $V_{\text{g}}$, as a function of the external magnetic field~$B$. The inset demonstrates $R(V_{\text{g}})$ curves at \myvareq{B}{}{0}{} and \myq{1.7}{T}. (b)~Log-log plot of the logarithm of $G_{\text{typ}}$ normalized over $G_0$ versus the magnetic field for the three edges. (c) Luttinger liquid parameter fits for the $3\|3\,\mathrm{\mu m}$ edge. (d)~Same as in~(b) but in the linear scale.}
\label{fig4}
\end{figure}

At \myq{0.5}{K}, introduction of the magnetic field results in the substantial edge conductance decrease and development of the strong mesoscopic fluctuations, see Fig.~\ref{fig4}(a). Here, the $I-V_{\text{sd}}$ characteristics are highly nonlinear, see Supplemental Material Fig.~7. In line with~\cite{PhysRevLett.123.056801}, such a behavior indicates the loss of topological protection under externally broken TRS. For the measurements of~$G_{\text{typ}}$ we focus on a narrow, \myq{20}{mV}, range of gate voltages in the CNP region so that the fluctuations are well reproducible in multiple $V_{\text{g}}$~sweeps, see Supplemental Material Fig.~8. Generally, we observe two ranges of magnetic field with two different exponents for the dependence~$l_{\text{B}}(B)$, see Fig.~\ref{fig4}(b). At low fields $B\lesssim0.1\,\text{T}$, the localization length~$l_{\text{B}}\propto B^{-\alpha}$, with $\alpha$ ranging approximately from $1.6$ to $2.8$. Despite this clear difference in the slope for different edges, the corresponding $K=3/2-1/\alpha$~\cite{PhysRevB.98.054205} lies in the range between $0.9$ and $1.1$, see Fig.~\ref{fig4}(c) with the data for the $3\|3\,\mathrm{\mu m}$ edge. This observation corresponds to the noninteracting disordered edge which should not localize with nonmagnetic disorder in the TRS case in contradiction with the observations of Fig.~\ref{fig3}. At higher fields up to $\approx$~\myq{0.5}{T} the dependence~$l_{\text{B}}(B)$ slows down, with $\alpha\approx1.1$ corresponding to $K\approx0.6$. Note that at relatively high fields the data follows the same slope for different edges, see also Supplemental Material Fig.~9 for additional data. The obtained value of~$K\approx0.6$ seemingly corresponds to a stronger inter-electron interaction which may reflect the renormalization of~$K$ with localization getting stronger. Still, $K\approx0.6$ is not sufficient enough to explain localization in the absence of magnetic field according to theoretical expectations for the disordered interacting edges~\cite{PhysRevLett.96.106401,PhysRevB.73.045322,PhysRevB.98.054205}.

We note that our observations of $\alpha\sim2$ at the relatively weak fields may be qualitatively expected also in the scenario of the edge coupled to random magnetic fluxes through the Fabry-Perot-type loops on the boundary of the 2D TI~\cite{PhysRevLett.109.246803}. Here, the inverse localization length~$l_{\text{B}}^{-1} \propto \ln{\left(G_{\text{typ}}(B)/G_0\right)}$ shows quadratic $B$-dependence at low $B$ followed by rapid growth and saturation when the magnetic flux through the average-area loop reaches $\approx 0.1\Phi_0=0.1h/e$, see Fig.~4 of~\cite{PhysRevLett.109.246803}. In our experiment, $B_{\text{sat}}\approx1\,\text{T}$, see Fig.~\ref{fig4}(d), corresponds to the average loop area of $20\,\text{nm}\times20\,\text{nm}$. 
While this estimate seems reasonable for the explanation of the $B$-driven localization, it does not explain the observed localization in the absence of external magnetic field.

In summary, we observed localization of the helical edge states in an \myq{8}{nm} HgTe quantum well in zero magnetic field at millikelvin temperatures. This observation is unlikely to be due to the magnetic disorder and the hyperfine interaction. While the most reasonable explanation is the interaction-driven localization of the dirty edges, the analysis of the magnetotransport data suggests that the strength of inter-electron interaction is insufficient to account for the observed localization. Our result possibly heralds the importance of many-body effects in the field of topological insulators where in terms of experiment interaction effects are poorly studied.

All the measurements were performed under the support of the Russian Science Foundation Grant No. 18-72-10135.
Fabrication of the devices was performed using the equipment of MIPT Shared Facilities Center.
We thank I.S.\,Burmistrov, Y.-Z.\,Chou, V.S.\,Khrapai, R.M.\,Nandkishore and L.\,Radzihovsky for useful discussions. We also thank E.M.\,Baeva, A.K.\,Grebenko, G.N.\,Goltsman, A.I.\,Kardakova and V.N.\,Zverev for technical assistance.
\widetext
\clearpage
\begin{center}
\textbf{\large Supplemental Material}
\end{center}
\setcounter{equation}{0}
\setcounter{figure}{0}
\setcounter{table}{0}
\setcounter{page}{1}
\makeatletter
\renewcommand{\theequation}{S\arabic{equation}}
\renewcommand{\figurename}{Supplemental Material Fig.}
\renewcommand{\bibnumfmt}[1]{[S#1]}
\renewcommand{\citenumfont}[1]{S#1}
\section*{Measurement details}
The filtering was performed using \myq{4}{nF}/\myq{0.2}{\mu H} $\pi$-filters at room temperature, followed by the N12-50F-257-0 flexible stainless steel coaxial cables, thermalized on the $4\text{K}$ and still plates, with total resistance of \myq{300}{Ohm} down to the coldplate. At coldplate, the coaxial cables were connected to the cinch connector on closed copper box which contained the chip. Inside the box, the sample was connected via twisted copper pairs additionally thermalized with a silver paste to a massive rod. Together with grounding~\myq{10}{nF} capacitors near the chip, the stainless coaxial cables also served as low-pass RC-filters. 

\section{Device Fabrication}
For patterning of the mesa we used e-beam lithography followed by wet mesa etching in an aqueous solution of KI:I$_2$:HBr. Further, Ti/Al contacts were evaporated right after \textit{in situ} cap removal with the Ar gun. The \myq{70}{nm} SiO$_2$ gate dielectric is then magnetron sputtered followed by Ti/Al gate electrode deposition. During fabrication the heating of the substrate was carefully controlled with the highest temperature of $80^\circ$C utilized for resist baking. 

\newpage
\section*{Nonlocal measurements}
\begin{figure}[h]
\begin{center}
\vspace{0mm}
\includegraphics[width=6in]{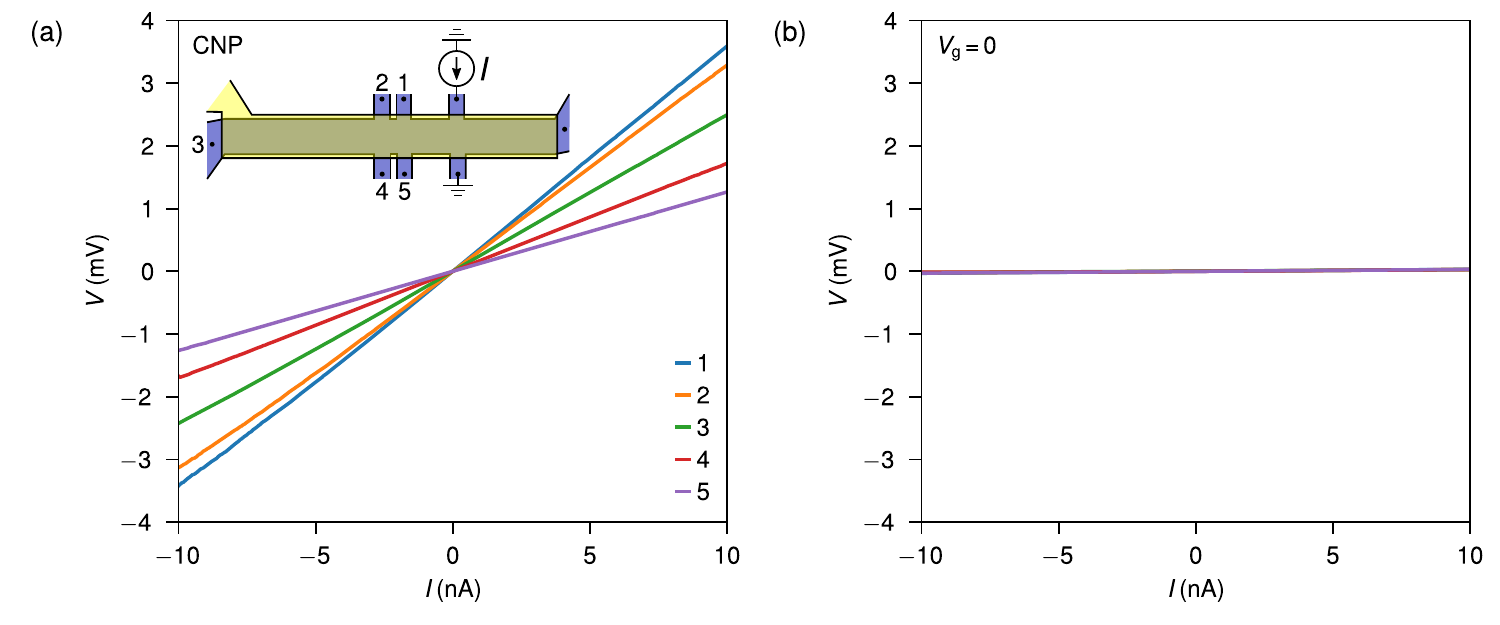}
\end{center}
\caption{Nonlocal transport measurements at~\myvareq{T}{}{4.2}{K}. Voltage probed at the terminals located aside from the bulk current path. (a)~\vg tuned to CNP region, so the bulk is insulating. Then terminals probe electrical potential of edge states, visualizing current flowing along the boundary of the sample. (b)~At \vgeq{0} 2D electron gas dominates the transport making voltage measured on terminals 1-5 negligible.}
\label{figSM1}
\end{figure}

\begin{figure}[h]
	\begin{center}
		\vspace{0mm}
		\includegraphics[width=3.1in]{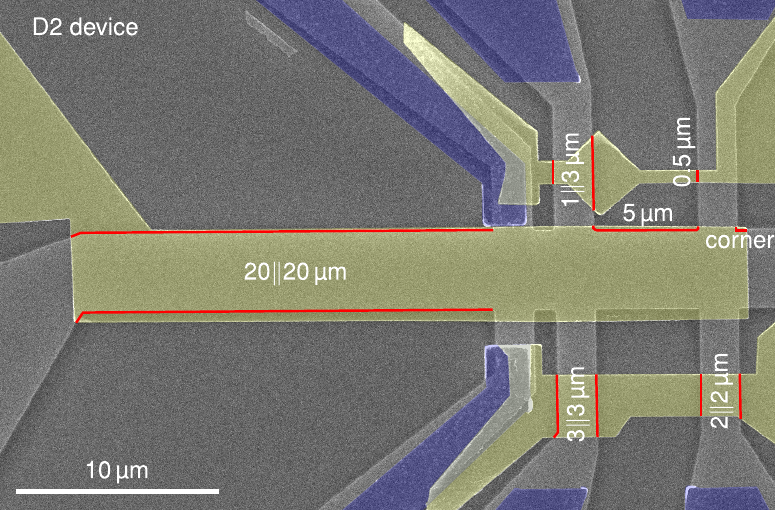}
	\end{center}
	\caption{False colored SEM micrograph of the device~D2. Studied edges are indicated by red lines captioned the same way they were mentioned in the text. }
	\label{figSM_edges}
\end{figure}
	
\clearpage
\section*{Localization}
\begin{figure}[h]
\begin{center}
\vspace{0mm}
\includegraphics[width=6in]{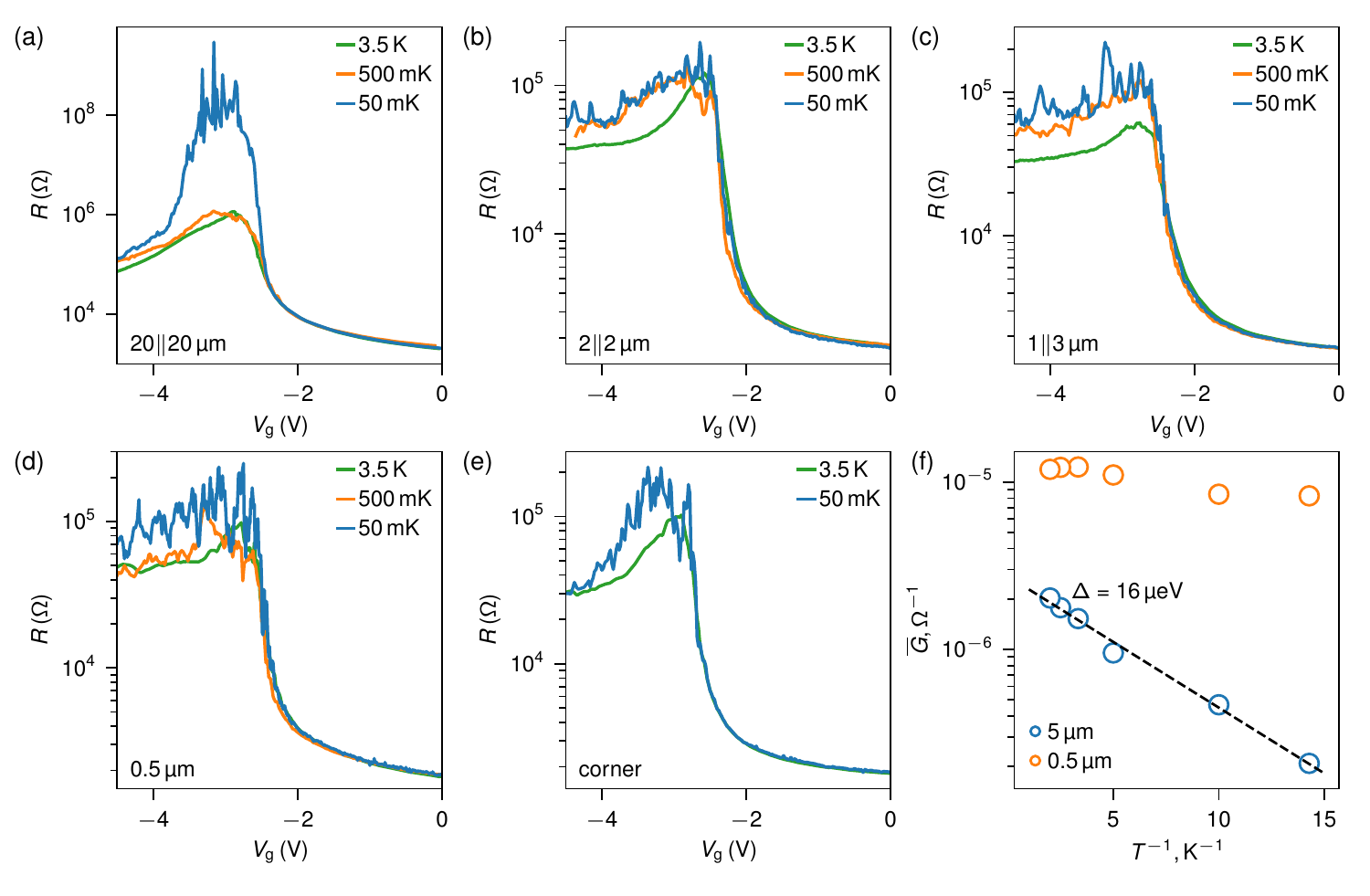}
\end{center}
\caption{Temperature dependence of edge transport. (a-e)~$R(V_\text{g})$ plots for five edges of different lengths. (f)~Temperature dependence of the mean conductance in the CNP region. For 5~$\mu$m edge black dashed line represents activation-like behavior with $\Delta=16~\mu\text{eV}$ exponent.}
\label{figSM2}
\end{figure}

\clearpage
\section*{Differential resistance}
\begin{figure}[h]
\begin{center}
\vspace{0mm}
\includegraphics[width=4.5in]{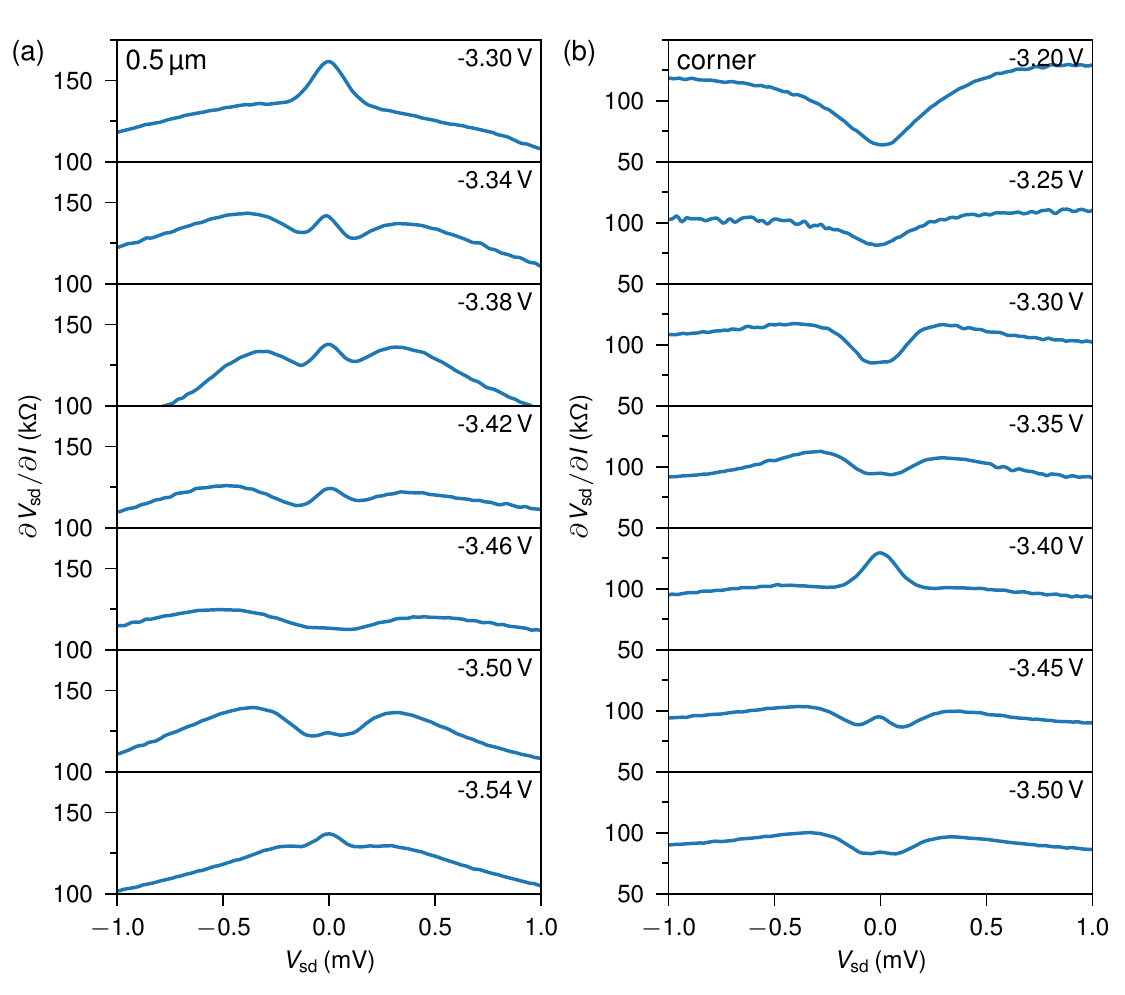}
\end{center}
\caption{Differential resistance of two short edges obtained at different \vg values inside CNP region at 0.5~K. (a)~\myq{0.5}{\mu m} and (b)~corner edges of D2 device.}
\label{figSMIV}
\end{figure}
\clearpage

\section*{Magnetotransport data}
\begin{figure}[h]
\begin{center}
\vspace{0mm}
\includegraphics[width=6.5in]{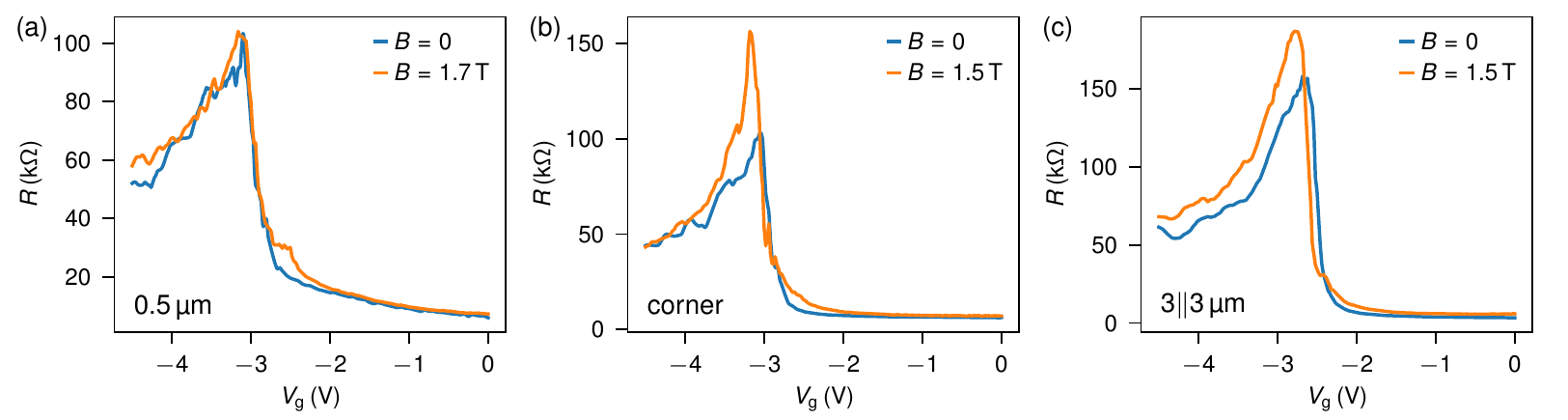}
\end{center}
\caption{Magnetic field dependence of the edge resistance at 4.2~K. (a)~\myq{0.5}{\mu m}, (b)~corner and (c)~$3\|3\,\mathrm{\mu m}$ edges of D2~device.}
\label{figSM7}
\end{figure}

\begin{figure}[h]
\begin{center}
\vspace{0mm}
\includegraphics[width=4in]{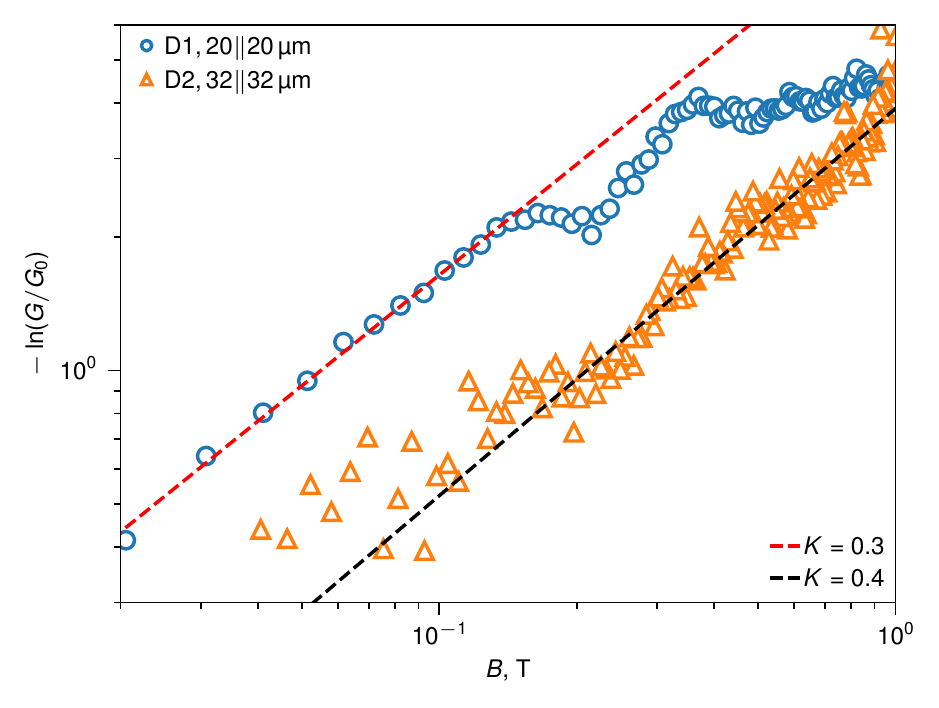}
\end{center}
\caption{Influence of the external magnetic field on the conductance of the long edges at~\myvareq{T}{}{0.5}{K}. Conductance of the $20\,\mathrm{\mu m}$ edge (device D2) and of the $32\,\mathrm{\mu m}$ edge (two-terminal device D3 processed similarly to~D1 and~D2) measured at fixed~$V_{\text{g}}$ in the CNP region, as a function of the external magnetic field~$B$.}
\label{figSMRG}
\end{figure}

\begin{figure}[h]
\begin{center}
\vspace{0mm}
\includegraphics[width=5in]{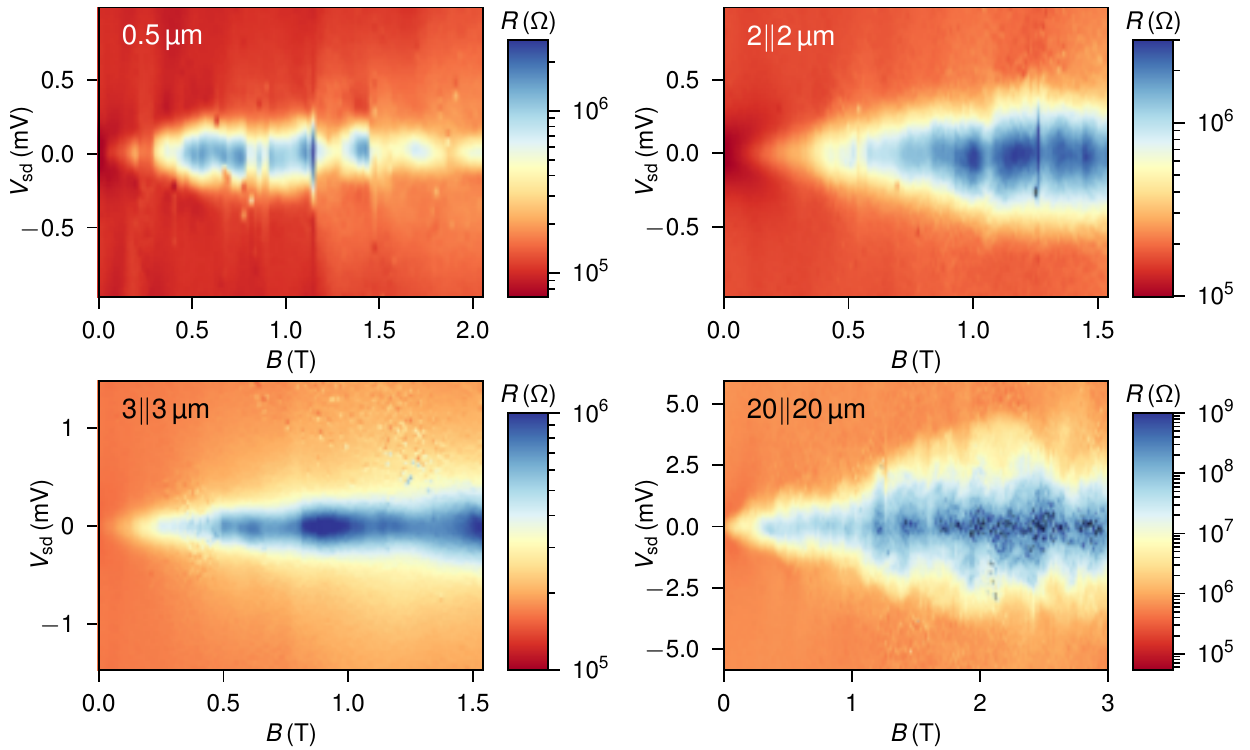}
\end{center}
\caption{Influence of the external magnetic field on the edge differential resistance at the CNP at~\myvareq{T}{}{0.5}{K}. Edges of different lengths of D2~device \myq{0.5}{\mu m}, $2\|2\,\mathrm{\mu m}$, $3\|3\,\mathrm{\mu m}$ and $20\|20\,\mathrm{\mu m}$  show qualitatively the same behavior with mobility gap opening on the order of hundreds $\mu\text{eV}$ (several mV) for shot (long) edges in $\sim 1$~T magnetic field.}
\label{figSM4}
\end{figure}

\begin{figure}[h]
\begin{center}
\vspace{0mm}
\includegraphics[width=4in]{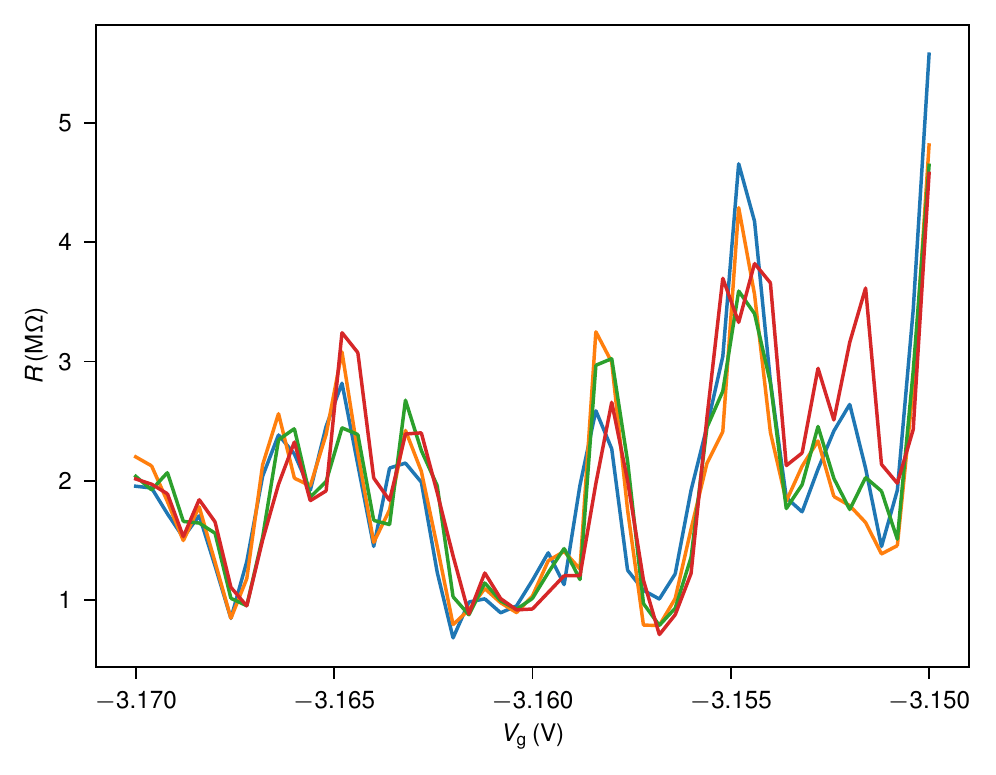}
\end{center}
\caption{Four consecutive $R(V_\text{g})$ measurements of \myq{0.5}{\mu m} edge of D2~device at 0.5~K in 1.7~T magnetic field. Mesoscopic oscillations are highly reproducible provided \vg sweep range is not too wide.}\label{figSM3}
\end{figure}

\begin{figure}[h]
\begin{center}
\vspace{0mm}
\includegraphics[width=5in]{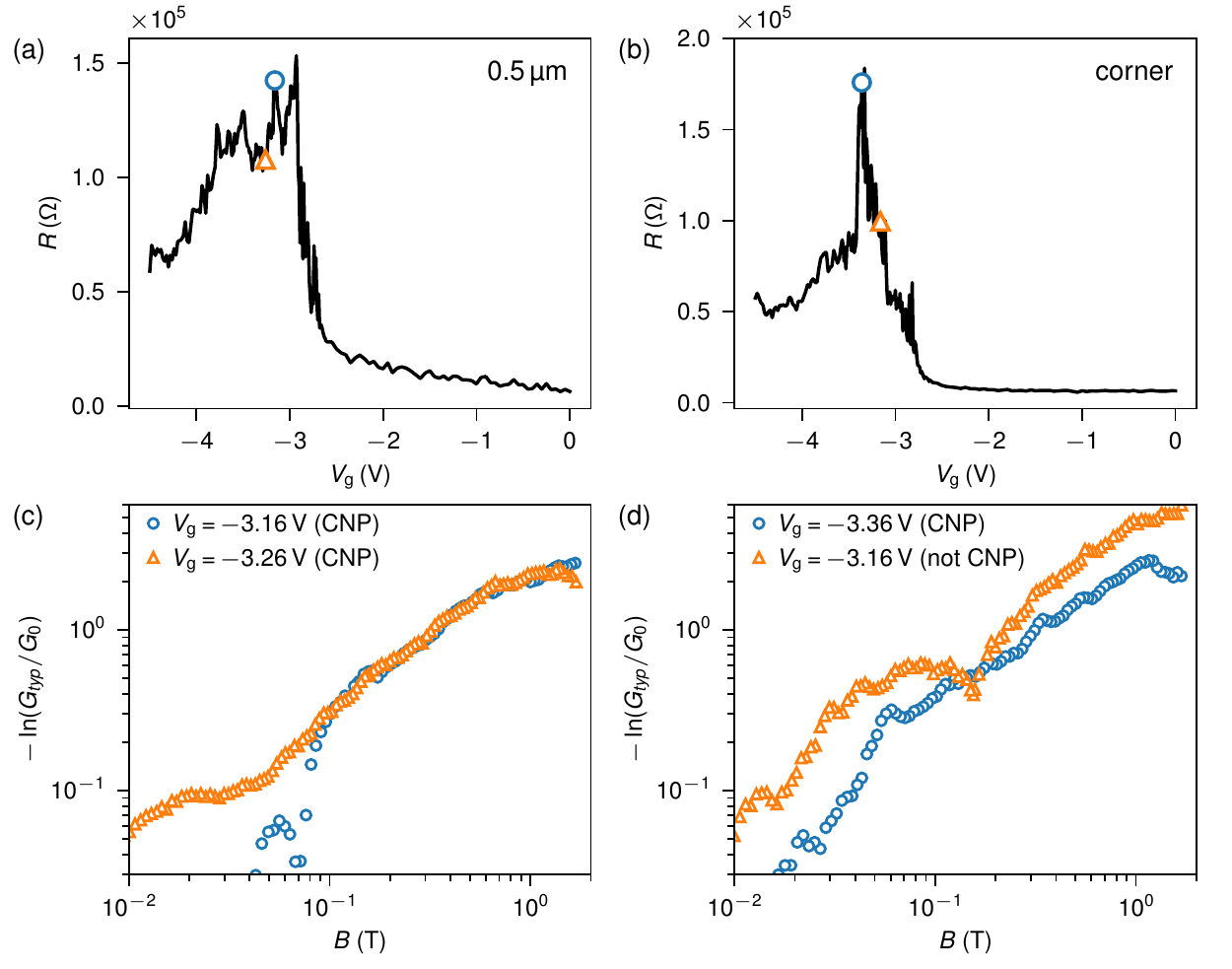}
\end{center}
\caption{Reproducibility of $G_{\text{typ}}(B)$ obtained at different \vg values. $R(V_\text{g})$ measurements at 0.5~K and without external magnetic field applied (a)~\myq{0.5}{\mu m} and (b)~corner edges of D2~device. (c)~Magnetotransport data of \myq{0.5}{\mu m} edge normalized over $G_0 \equiv G_{\text{typ}}(0)$. Symbols on~(a) denote position respective to $R(V_\text{g})$ curve at which $G_{\text{typ}}(B)$ was measured. While low-$B$ data differs due to its strong sensitivity to measurement accuracy of $G_0$, high-$B$ data coincides for two \vg values. (d)~The similar data for corner edge. Here the importance of proper choice of \vg is clear: CNP should determined by the linear-response resistance peak in $B=0$ $G_{\text{typ}}^{-3.16}(0) > G_{\text{typ}}^{-3.36}(0)$, while in $B\gg0$ situation can be changed $G_{\text{typ}}^{-3.16}(1.7~\text{T}) \ll G_{\text{typ}}^{-3.36}(1.7~\text{T})$ ($4.5\cdot10^{-8}$~$\Omega^{-1}$ and $1.2\cdot10^{-6}$~$\Omega^{-1}$ respectively).}
\label{figSM5}
\end{figure}
\end{document}